\documentclass[aps,prl,twocolumn,10pt,groupedaddress,showpacs]{revtex4-1}

\usepackage{graphicx}
\usepackage{amssymb}
\usepackage{amsthm}
\usepackage{amsfonts,amsmath,latexsym,bbm}
\usepackage{ae,aecompl}                          
\usepackage{dsfont}
\usepackage{tensor}
\usepackage{color}
\usepackage{tikz-cd}
\usepackage{tikz}
\usetikzlibrary{decorations.markings}
\usetikzlibrary{matrix,arrows}
\usepackage{natbib}
\usepackage{upgreek}
\usepackage{hyperref}

\newcommand{\de}{\mathrm d}

\newcommand{\tsr}[1]{\overset\leftrightarrow{#1}}
\newcommand{\ext}{_{\rm ext}}
\newcommand{\tot}{_{\rm tot}}
\newcommand{\ind}{_{\rm ind}}

\renewcommand{\vec}[1]{\boldsymbol{#1}}
\newcommand{\h}{\hspace{1pt}}

\definecolor{darkred}{rgb}{0.8,0,0}

\newcommand{\hh}{\hspace{0.3pt}}

\renewcommand{\i}{\mathrm i}

\DeclareMathAlphabet{\mathbbmsl}{U}{bbm}{m}{sl}

\begin{document}


\title{Microscopic theory of refractive index applied to metamaterials: \\ Effective current response tensor corresponding to standard relation $n^2 =\varepsilon_{\rm eff} \h \mu_{\rm eff}$}


\author{G.\,A.\,H.~Schober}
\email[]{schober@physik.rwth-aachen.de}
\affiliation{Institute for Theoretical Solid State Physics, RWTH Aachen University, Otto-Blumenthal-Stra\ss e, 52074 Aachen}
\author{R.~Starke}
\email[]{Ronald.Starke@physik.tu-freiberg.de}
\affiliation{Institute for Theoretical Physics, TU Bergakademie Freiberg, Leipziger Stra\ss e 23, 09596 Freiberg, Germany}


\date{\today}

\begin{abstract}
In this article, we first derive the {\it wavevector- and frequency-dependent, microscopic current response tensor} which corresponds to the ``macroscopic'' ansatz $\vec D = \varepsilon_0\varepsilon_{\rm eff}\vec E$ and $\vec B= \mu_0\mu_{\rm eff}\vec H$ with wavevector- and frequency-{\itshape independent}, ``effective'' material constants $\varepsilon_{\rm eff}$ and $\mu_{\rm eff}$. 
We then deduce the electromagnetic and optical properties of this effective material model by employing exact, microscopic response relations. In particular,
we argue that for recovering the standard relation $n^2=\varepsilon_{\rm eff}\hh \mu_{\rm eff}$ between the refractive index and the effective material constants, it is imperative to start from the microscopic wave equation in 
terms of the {\it transverse dielectric function}, $\varepsilon_{\rm T}(\vec k,\omega)=0$. 
On the phenomenological side, our result is especially relevant for metamaterials research, which draws directly on the standard relation for the refractive index in terms of effective material constants. Since for a wide class of materials the current response tensor can be calculated from first principles and compared to the model expression derived here, this work also paves the way for a systematic search for new metamaterials.
\end{abstract}


\maketitle

\section{Introduction}

The traditional approach to electrodynamics in media \cite{Jackson,Griffiths,Landau} is based on the division of electric charge and current densities into
``free'' and ``bound'' contributions, combined with the so-called ``macroscopic'' Maxwell equations, which are usually written in the form
\begin{align}
 \nabla\cdot\vec D & = \rho_{\rm f} \,,  		\label{eq_MaxMac1}\\[2pt]
 \nabla\times\vec E &= -\partial_t\vec B \,, 		\label{eq_MaxMac2}\\[2pt]
 \nabla\cdot\vec B &= 0 \,, 				\label{eq_MaxMac3}\\[2pt]
 \nabla\times\vec H &= \vec j_{\rm f} + \partial_t\vec D \,.\label{eq_MaxMac4}
\end{align}
These equations have to be complemented by the so-called ``constitutive laws'', which are---more often than not---assumed to be simple
linear relations between $\vec D$ and $\vec E$ as well as $\vec H$ and $\vec B$, i.e.,
\begin{align}
 \vec D & = \varepsilon_0 \h \varepsilon_{\rm r} \h \vec E  \,, 		\label{eq_permitt}\\[2pt]
 \vec H & = \mu_0^{-1} \mu_{\rm r}^{-1} \vec B \,,			\label{eq_permea}
\end{align}
where $\varepsilon_{\rm r}$ and $\mu_{\rm r}$ are the {\itshape relative} permittivity and permeability, respectively. To lighten the notation, we will in the following suppress the subscript $\rm r$ and simply write $\varepsilon \equiv \varepsilon_{\rm r}$ and $\mu \equiv \mu_{\rm r}$ for these dimensionless quantities. (They should, however, not be confused with the corresponding {\itshape absolute} quantities given by $\varepsilon_0 \hh \varepsilon_{\rm r}$ and $\mu_0 \hh \mu_{\rm r}$.) In particular, the fundamental field equations \eqref{eq_MaxMac1}--\eqref{eq_MaxMac4} determine the involved field quantities $\{ \vec D, \vec H, \vec E, \vec B\}$ only once the constitutive laws are specified. 
Without these the fields would remain underdetermined. The constitutive laws on their side are usually assumed to be given empirically.
In the simplest case, one thinks of them as being formulated in terms of {\itshape effective} material {\it constants}, which are measurable in principle.
In general, though, the constitutive laws could be much more complicated (involving retardation effects, non-linearities, etc.). Consequently, the field equation 
for, say, the divergence of the ``magnetic field'' $\vec H$ is material dependent in the Standard Approach described above. In principle, 
this divergence has to be determined by plugging
the relation $\vec B=\vec B[\vec H]$ into the field equation~\eqref{eq_MaxMac3}.

However, along with the advent of {\it ab initio materials physics} \cite{Giuliani,Kohanoff,Martin},
a new, {\it microscopic} approach to electrodynamics in media has been developed \cite{NozieresPines,NozieresPines2,KeldyshKirzhnitz,Kittel,Fliessbach,Melrose1Book,Melrose2Book}.
Its basis is the division of both the electromagnetic source terms (i.e., charge and current densitites) and the electromagnetic fields into their respective
{\it external} and {\it induced} contributions \cite{Bruus,MartinRothen,Kaxiras,SchafWegener,Hanke,Strinati}, 
where the term ``induced'' means ``generated under the influence of {\it externally applied} fields''.
For convenience, one also considers ``total'' fields, which are defined as the sum of external and induced contributions.
In this microscopic approach, all electric and magnetic fields are uniquely determined by the microscopic Maxwell equations \cite{Zangwill},
\begin{align}
 \nabla\cdot\vec E(\vec x, t) 	&= \rho(\vec x, t)/\varepsilon_0 \,, 	\\[2pt]
 \nabla\times\vec E(\vec x, t)	&= -\partial_t\vec B(\vec x, t) \,, 	\\[2pt]
 \nabla\cdot\vec B(\vec x, t)	&= 0 \,, 			\\[2pt]
 \nabla\times\vec B(\vec x, t)	&= \mu_0\h\vec j(\vec x, t) + \varepsilon_0\mu_0\h\partial_t\vec E(\vec x, t) \,.
\end{align}
These equations now hold separately for the external, induced and total fields in terms of the respective external, induced and total sources.
Therefore, all electric and magnetic fields are uniquely determined {\it independently of the material under consideration}. 
The latter's influence only comes into play when we  consider {\it response relations}, which mostly express an induced field quantity in terms of an externally
applied field via a corresponding {\it response function}. 
In principle, these microscopic response functions can be calculated from {\it first-principles} (many-body Schr\"{o}dinger equation combined with Kubo formalism).
Hence, they do {\it not} constitute freely adjustable material parameters. Instead, they can be predicted theoretically as~well.

An important quantity in the {\itshape ab initio} context is given, for example, by the {\itshape density response function} $\upchi$\h, which is implicitly defined by
\begin{equation} \label{def_density_response}
 \rho\ind(\vec x,t) = \int \! \de^3 \vec x' \int \! c \, \de t'\, \upchi(\vec x,\vec x';t-t') \h \varphi\ext(\vec x',t')\,,
\end{equation}
where $\varphi\ext$ denotes the externally applied scalar potential. In particular, in the microscopic approach, response functions are in general
given in terms of non-local (possibly tensorial) integral kernels. Only for homogeneous systems (which, admittedly, constitute the most
important practical application in theoretical materials science), the response functions depend only on the coordinate difference, such that the
response laws have a purely multiplicative structure {\itshape in Fourier space}, i.e.,
\begin{equation}
 \rho\ind(\vec k,\omega) = \upchi(\vec k,\omega) \h \varphi\ext(\vec k,\omega) \,.
\end{equation}
Once the microscopic density response function is given, a more traditional material property like the (relative) {\itshape dielectric function} can be calculated
by means of the standard relation \cite[\S\,5.1]{EDWave}
\begin{equation}
 \varepsilon^{-1}(\vec k,\omega) = 1 + \frac{\upchi(\vec k,\omega)}{\varepsilon_0|\vec k|^2} \,.
\end{equation}
In principle, this quantity can be identified with the {\it permittivity} used in the Standard Approach (see Eq.~\eqref{eq_permitt}).
This has to be shown on the basis of the Fundamental Field Identifications given by \cite{ED1,ED2}
\begin{align}
 \vec D(\vec x, t) &= \varepsilon_0 \hh \vec E\ext(\vec x, t) \,,\\[1pt]
 \vec P(\vec x, t) &= -\varepsilon_0 \hh \vec E\ind(\vec x, t) \,, \\[1pt]
 \vec E(\vec x, t) &= \vec E\tot(\vec x,t) \,,
\end{align}
and by
\begin{align}
\mu_0 \hh \vec H(\vec x, t) &= \vec B\ext(\vec x, t) \,, \\[1pt]
\mu_0 \hh \vec M(\vec x, t) &= \vec B\ind(\vec x, t) \,, \\[1pt]
\vec B(\vec x, t) &= \vec B\tot(\vec x, t) \,.
\end{align}
As a matter of principle, these identifications relate the microscopic fields used in {\itshape ab initio} materials science
to their macroscopic counterparts used in the Standard Approach.

However, the Fundamental Field Identifications lead to the following problem
which does not exist in the Standard Approach: if all electromagnetic fields (external, induced and total) are already completely determined
by means of their respective Maxwell equations, while on the other hand the induced fields are also 
determined in terms of the external fields via the corresponding ``direct'' response functions (or in terms
of the total fields via the ``proper'' response functions \cite[\S\,2.3]{Refr}), then apparently there exists an {\itshape over\-{}determination}
in the theory which could in principle lead to contradictions. For example, in the traditional approach the expression \mbox{$\nabla\cdot\vec H$} simply
remains undetermined, while in the microscopic approach we necessarily obtain $\nabla\cdot\vec H=0$ on the basis of the Fundamental
Field Identifications, although at the same time we have $\vec B=\mu\vec H$ (within the limits of linear response theory, where $\mu$ in general denotes
a tensorial integral kernel). 
The resolution of this apparent paradox lies in another surprising feature of the microscopic approach, which 
distinguishes it sharply from its traditional macroscopic counterpart: the microscopic response functions cannot be prescribed arbitrarily.
Instead, they are subject to {\it constraints} which guarantee the validity of the microscopic Maxwell equations.
In particular, this also implies that the response functions are in general not independent of each other, but interrelated
by the Universal Response Relations \cite{ED1}. Concretely, it turns out that the microscopic {\it current response tensor} determines all other
linear electromagnetic response properties uniquely and explicitly \cite{ED1,ED2}. Thus, the Universal Response Relations greatly
facilitate both theoretical calculations and experimental measurements as they allow for the deduction of one response function from another.

Conceptually, however, the somewhat shocking implication of this microscopic approach is that the standard formula for the refractive index in terms of the relative permittivity and permeability,
\begin{equation}
 n^2 (\omega) \stackrel{?}{=} \varepsilon(\omega) \h \mu(\omega) \label{eq_MaxwellStand} \,,
\end{equation}
cannot be true {\itshape in this form} \cite{EDWave,Refr,EDFresnel}, {\itshape i.e., as a formula expressing the refractive index in terms of response functions}. 
Instead, its allegedly approximate version,
\begin{equation}
 n^2 (\omega) = \varepsilon(\omega) \,, \label{eq_MaxwellMoreCorrect}
\end{equation}
turns out to be the more correct formula, which can be justified microscopically \cite{Refr}. Here, it is understood that the involved frequency-dependent quantities
refer to the {\itshape macroscopic limit} ($\vec k\rightarrow \vec 0$) of microscopic response functions, which can be calculated from first principles.

Fortunately, in most cases the failure of the standard relation \eqref{eq_MaxwellStand} does not pose any serious problems \cite{EDFresnel}.
In fact, textbooks in condensed matter theory often even {\it define} \cite{Kittel,Ashcroft,Cardona} the refractive index by the allegedly approximate relation \eqref{eq_MaxwellMoreCorrect}.
Furthermore, a {\it bulk material} where the standard formula \eqref{eq_MaxwellStand} would apply with independently obtained material
parameters $\varepsilon$, $\mu$ and $n$ is not known.
In the research domain of so-called {\it metamaterials}, however, one draws directly on the original Maxwell relation \eqref{eq_MaxwellStand}
if  only with ``effective'' (i.e., not calculated from first principles) {\it material parameters} (not response functions) $\varepsilon_{\rm eff}(\omega)$
and $\mu_{\rm eff}(\omega)$ \cite{Pendry}. 
Concretely, it has been argued by V.~Veselago that $n$ should be regarded as a negative number if both $\varepsilon_{\rm eff} < 0$ and $\mu_{\rm eff} < 0$ \cite{Veselago}. 
Such a negative effective permeability can occur in {\it artificial} materials by exploiting the concept of a split-ring resonator \cite{Pendry, Smith00}. 
An anomalous light refraction at metamaterials has been observed experimentally \cite{Shelby}. Therefore, 
metamaterials are regarded as promising candidates for technological applications such as {\it superlenses} \cite{PendrySuper,SmithPendry} and {\it invisibility cloaks} \cite{Ergin}.
For the longest time, this line of research has been pursued in complete independence from first-principles materials science.

However, in their recent groundbreaking work \cite{Forcella}, D.\ Forcella {\itshape et~al.} have---apparently for the first time---approached metamaterials theory from the {\itshape ab initio} point of view.
Concretely, they have stressed the fact that within first-principles materials science, the complete information about all linear electromagnetic
response properties (including the refractive index) can alternatively be encoded in the microscopic, 
non-local dielectric tensor (see Refs.~\cite{Forcella} or \cite{KeldyshKirzhnitz,ED1}), and in fact, their profound analysis
is based on this quantity {\it exclusively}. Unfortunately, the drawback is that the original argument by V.~Veselago is 
then not applicable in its na\"{i}ve form anymore due to the lack of effective ``electric'' and ``magnetic'' material parameters, the product of which could be considered
as an expression for the refractive index as in Eq.~\eqref{eq_MaxwellStand}. Hence, the question arises whether it could even 
be possible to reproduce Veselago's original argument within the framework
of microscopic electrodynamics in media \cite{ED1,ED2,Refr} under suitable assumptions which then, admittedly, would have to be more specific than the general approach by \mbox{D.~Forcella {\itshape et.~al.}}

With this state of affairs, we actually face two fundamental questions:
\begin{enumerate}
 \item Although the refractive index is microscopically not determined by the standard formula \eqref{eq_MaxwellStand}, 
 is it still possible to have a material whose microscopic response functions involve two (constant) {\it material parameters,} 
 which have an interpretation as ``effective'' electric permittivity and permeability, such that the standard formula \eqref{eq_MaxwellStand}
 instead holds in terms of these ``effective'' material constants?
 \item More generally, to which {\it current response tensor} does the simple macroscopic ansatz \eqref{eq_permitt}--\eqref{eq_permea}  correspond, which is 
 used in the traditional approach for the derivation of the standard formula \eqref{eq_MaxwellStand}?
\end{enumerate}
To answer these questions is precisely the aim of the present~article.

\section{Phenomenological derivation}

We start from the ``macroscopic'' Maxwell equations written in Fourier space as
\begin{align}
 \vec k \cdot \vec B(\vec k, \omega) & = 0 \,, \\[3pt]
 \vec k \times \vec E(\vec k, \omega) - \omega \vec B(\vec k, \omega) & = 0 \,, \\[3pt]
 \i \vec k \cdot \vec D(\vec k, \omega) & = \rho_{\rm f}(\vec k, \omega) \,, \\[3pt]
 \i\vec k \times \vec H(\vec k, \omega) + \i\omega \vec D(\vec k, \omega) & = \vec j_{\rm f}(\vec k, \omega)  \,.
\end{align}
By means of the first two, homogenous equations, we can introduce the potentials
\begin{align}
 \vec B(\vec k, \omega) & = \i\vec k \times \vec A(\vec k, \omega) \,, \\[3pt]
 \vec E(\vec k, \omega) & = -\i\vec k \h \varphi(\vec k, \omega) + \i \omega \vec A(\vec k, \omega) \,.
\end{align}
Furthermore, in the last two, inhomogeneous Maxwell equations, we substitute
\begin{align}
 \vec D(\vec k, \omega) & = \varepsilon_0 \h \varepsilon_{\rm eff} \h \vec E(\vec k, \omega)  \,, \\[3pt]
 \vec H(\vec k, \omega) & = \mu_0^{-1} \mu^{-1}_{\rm eff} \h \vec B(\vec k, \omega) \,,
\end{align}
with the {\itshape effective} permittivity and permeability, $\varepsilon_{\rm eff}$ and $\mu_{\rm eff}$\hh, which are assumed 
to be (wavevector- and fre\-{}quency-independent) {\itshape constants.} 
We note in passing that in principle, these material constants could {\itshape ex post} be promoted to frequency-dependent functions,
$\varepsilon_{\rm eff}(\omega)$ and $\mu_{\rm eff}(\omega)$, in order to include, for instance, the effects of dissipation
without modifying the central results of the present article. In any case, we thus obtain the inhomogeneous equations for the potentials (suppressing $\vec k$ and $\omega$ dependencies in the notation),
%
\begin{equation}
 \varepsilon_0 \hh \varepsilon_{\rm eff} \h (|\vec k|^2 \h \varphi - \omega \h \vec k \cdot \vec A) = \rho_{\rm f} \,,
\end{equation}
 and
\begin{equation}
 \frac{1}{\mu_0 \hh \mu_{\rm eff}} \h \big(|\vec k|^2 \vec A - \vec k \h (\vec k \cdot \vec A) \big) + \varepsilon_0 \hh \varepsilon_{\rm eff}\h (\omega \hh \vec k \h \varphi - \omega^2 \vec A) = \vec j_{\rm f} \,.
\end{equation}
In matrix form, these equations can be rewritten in terms of Lorentz four-vectors as
%
%
%
%
\begin{equation} \label{zwischen_1}
\begin{aligned}
  \mu_0 \h \bigg( \!\! \begin{array}{c} c \rho_{\rm f} \\[3pt] \vec j_{\rm f} \end{array} \!\! \bigg) & = \varepsilon_{\rm eff} \, 
  \Bigg( \! \begin{array}{cc} |\vec k|^2 & -\frac{\omega}{c} \h \vec k^{\rm T} \\[3pt] \frac{\omega}{c} \h \vec k & -\frac{\omega^2}{c^2} \end{array} \! \Bigg) 
  \bigg( \!\! \begin{array}{c} \varphi/c \\[3pt] \vec A \end{array}\!\!  \bigg) \\[3pt]
  & \quad \, + \frac{1}{\mu_{\rm eff}} \h \bigg( \! \begin{array}{cc} 0 & 0 \\[3pt] 0 & |\vec k|^2 - \vec k \vec k^{\rm T} \end{array}\!\!  \bigg) \bigg( \!\! \begin{array}{c} \varphi/c \\[3pt] \vec A \end{array}\!\!  \bigg) \,.
\end{aligned}
\end{equation}
Finally, defining the $(4 \times 4)$ matrices
\begin{align}
 M_{\rm e}(\vec k, \omega) & \stackrel{\rm def}{=} \Bigg( \! \begin{array}{cc} |\vec k|^2 & -\frac{\omega}{c} \h \vec k^{\rm T} \\[3pt] \frac{\omega}{c} \h \vec k & -\frac{\omega^2}{c^2} \end{array} \! \Bigg) \,, \\[3pt]
 M_{\rm m}(\vec k, \omega) & \stackrel{\rm def}{=} \bigg( \! \begin{array}{cc} 0 & 0 \\[3pt] 0 & |\vec k|^2 - \vec k \vec k^{\rm T} \end{array}\!\!  \bigg) \,,
\end{align}
we can rewrite Eq.~\eqref{zwischen_1} compactly as
\begin{equation} \label{compact_result}
 \mu_0 \h j_{\rm f} = \bigg(\varepsilon_{\rm eff} \h M_{\rm e} + \frac{1}{\mu_{\rm eff}} \h M_{\rm m} \bigg) A \,,
\end{equation}
where $j \equiv j^\mu = (c\rho, \hh \vec j)^{\rm T}$ and $A^\nu = (\varphi/c, \vec A)$ are the four-current and four-potential, respectively.
In the following, this relation will form the basis for the identification of the {\itshape fundamental response tensor} \cite[\S\,5.1]{ED1} in this effective model for metamaterials.

\section{Proper fundamental response tensor}

In order to perform the transition from the traditional macroscopic approach to electrodynamics in media to its modern microscopic counterpart,
we now identify $j_{\rm f} \equiv j_{\rm ext} $ and  $A \equiv A_{\rm tot}$ \cite{ED1,ED2} such that Eq.~\eqref{compact_result} turns into
\begin{equation} \label{jext}
 \mu_0 \h j_{\rm ext} = \bigg(\varepsilon_{\rm eff} \h M_{\rm e} + \frac{1}{\mu_{\rm eff}} \h M_{\rm m} \bigg) A_{\rm tot} \,.
\end{equation}
Next, we use that \cite[Eq.~(3.30)]{ED1}
\begin{equation}
 \mu_0 \h j_{\rm tot}(k) = k^2 P_{\rm T}(k) \h A_{\rm tot}(k) \,,
\end{equation}
where $k^\mu=(\omega/c, \h \vec k)^{\rm T}$ denotes the four-momentum, $k^2 = k^\mu k_\mu = |\vec k|^2 - \omega^2/c^2$, and the {\itshape Minkowskian transverse projector} is given by \cite[\S\,2.1 and \S\,2.2]{EDFullGF}
\begin{equation}
 P_{\rm T}(\vec k, \omega)= \frac{1}{|\vec k|^2 - \omega^2 /c^2} \, \Bigg( \! \begin{array}{cc} |\vec k|^2 & -\frac{\omega}{c} \h \vec k^{\rm T} \nonumber \\[5pt] \frac{\omega}{c} \h \vec k & |\vec k|^2 - \frac{\omega^2}{c^2} - \vec k \vec k^{\rm T} \end{array} \! \Bigg) \,,
 \end{equation}
 which is equivalent to
 \begin{equation}
 P_{\rm T}(k) = \frac{1}{k^2} \h \big(M_{\rm e}(k) + M_{\rm m}(k) \big)\,.
\end{equation}
Together, these formulae imply the identity
\begin{equation}
 \mu_0 \h j_{\rm tot} = (M_{\rm e} + M_{\rm m}) \h A_{\rm tot} \,.
\end{equation}
Combining this with Eq.~\eqref{jext} yields
\begin{align}
 \mu_0 \h j_{\rm ind} & = \mu_0 \h j_{\rm tot} - \mu_0 \h j_{\rm ext} \\[2pt]
 & = \bigg((1-\varepsilon_{\rm eff}) \h M_{\rm e} + \bigg( 1 - \frac{1}{\mu_{\rm eff}}\bigg) M_{\rm m} \bigg) \h A_{\rm tot} \,.
\end{align}
We now interpret the term in brackets as the {\itshape proper} fundamental response tensor \cite{ED1,Refr,EDOhm}, which is hence given by
\begin{equation} \label{proper_fund}
 \mu_0 \, \widetilde \chi(\vec k, \omega) = (1-\varepsilon_{\rm eff}) \h M_{\rm e}(\vec k, \omega) 
 + \bigg( 1 - \frac{1}{\mu_{\rm eff}}\bigg) M_{\rm m}(\vec k, \omega) \,.
\end{equation}
One easily checks that $M_{\rm e}$ and $M_{\rm m}$ separately fulfill the constraints
\begin{equation}
 k_\mu \h M\indices{^\mu_\nu}(k) = M\indices{^\mu_\nu}(k) \h k^\nu = 0\,,
\end{equation}
and thus they are completely determined by their respective spatial parts,
\begin{align}
 \tsr M_{\rm e}(\vec k, \omega) & = -\frac{\omega^2}{c^2} \h \tsr 1 \,, \\[3pt]
 \tsr M_{\rm m}(\vec k, \omega) & = |\vec k|^2 \h \tsr 1 - \vec k \vec k^{\rm T} = |\vec k|^2 \h \tsr P_{\rm T}(\vec k) \,,
\end{align}
where $\tsr P_{\rm T}(\vec k)$ denotes the {\itshape Cartesian transverse projector} \cite[\S\,2.1]{EDWave}.
The proper fundamental response tensor \eqref{proper_fund} therefore fulfills the same constraints, and it is completely determined by the {\itshape proper current response tensor,}
\begin{equation}
 \mu_0 \h \tsr{\widetilde \chi}(\vec k, \omega) = (\varepsilon_{\rm eff} - 1) \, \frac{\omega^2}{c^2} \h \tsr 1 + \bigg( 1 - \frac{1}{\mu_{\rm eff}}\bigg) \h |\vec k|^2 \h \tsr P_{\rm T}(\vec k) \,.
\end{equation}
We can write this equivalently as \cite[Appendix D.1]{EffWW}
\begin{equation}\label{eq_IsoForm}
 \tsr{\widetilde \chi}(\vec k, \omega) = \widetilde \chi_{\rm L}(\vec k, \omega) \h \tsr P_{\rm L}(\vec k) + \widetilde \chi_{\rm T}(\vec k, \omega) \h \tsr P_{\rm T}(\vec k) \,,
\end{equation}
with the {\itshape longitudinal} and {\itshape transverse} proper current response functions
\begin{align}
 \widetilde \chi_{\rm L}(\vec k, \omega) & = \varepsilon_0 \h (\varepsilon_{\rm eff} - 1) \, \omega^2 \,, \label{propL} \\[3pt]
 \widetilde \chi_{\rm T}(\vec k, \omega) & = \varepsilon_0 \h (\varepsilon_{\rm eff} - 1) \, \omega^2 + \frac{1}{\mu_0} \h \bigg( 1 - \frac{1}{\mu_{\rm eff}}\bigg) \h |\vec k|^2 \,. \label{propT}
\end{align}
In particular, this shows that the phenomenological model defined by the fundamental response tensor \eqref{proper_fund} describes a homogeneous and isotropic system.

Thus, Eq.~\eqref{proper_fund} represents the first central result of this article. It gives  the {\it microscopic, wavevector- and frequency-dependent (proper) fundamental response tensor}  which corresponds
to the traditional ansatz defined by Eqs.~\eqref{eq_permitt}--\eqref{eq_permea}.
In fact, this microscopic response tensor depends on only two ``effective'' material constants, $\varepsilon_{\rm eff}$ and $\mu_{\rm eff}$\hh. 

However, it remains to show that: (i) although these {\it material constants}
do not coincide with the electric permittivity and the magnetic permeability (in the sense of {\it response functions}), 
they can still be interpreted as their ``effective''
versions; (ii) the microscopic wave equation leads to a refractive index which is simply given by the product of these effective material parameters, hence $n^2 = \varepsilon_{\rm eff} \h \mu_{\rm eff}$\hh.

\section{Effective permittivity and permeability}

For an isotropic system, the dielectric tensor has an analogous form as Eq.~\eqref{eq_IsoForm}.
The resulting longitudinal and transverse dielectric functions are related to the \linebreak

\pagebreak \noindent
corresponding proper current response functions by \cite[\S\,5.1]{EDWave}
\begin{align}
 \varepsilon_{\rm L}(\vec k, \omega) & = 1 + \frac{1}{\varepsilon_0 \h \omega^2} \, \widetilde \chi_{\rm L}(\vec k, \omega) \,, \\[3pt]
 \varepsilon_{\rm T}(\vec k, \omega) & = 1 + \frac{1}{\varepsilon_0 \h (\omega^2 - c^2 |\vec k|^2)} \, \widetilde \chi_{\rm T}(\vec k, \omega) \,.
\end{align}
Using Eqs.~\eqref{propL}--\eqref{propT}, we therefore obtain
\begin{align}
 \varepsilon_{\rm L}(\vec k, \omega) & = \varepsilon_{\rm eff} \,, \\[3pt]
 \varepsilon_{\rm T}(\vec k, \omega) & = \frac{\varepsilon_{\rm eff} \, \omega^2 - \mu_{\rm eff}^{-1} \h c^2 |\vec k|^2}{\omega^2 - c^2 |\vec k|^2} \,. \label{epsT}
\end{align}
Hence, both the longitudinal and the transverse dielectric function fulfill
\begin{equation}
 \lim_{|\vec k|\rightarrow 0}\varepsilon_{\rm L/T}(\vec k, \omega) = \varepsilon_{\rm eff}\,,
\end{equation}
and thus $\varepsilon_{\rm eff}$ can indeed be interpreted as an ``effective'' permittivity.
In particular, the longitudinal dielectric function is even {\itshape constant} and given by $\varepsilon_{\rm eff}$\hh. 
We remark, however, that this does not imply a proportionality between the external and the total electric field. Instead, 
we have the following relations between the longitudinal and transverse components of the respective fields:
\begin{align}
 \vec E_{\rm ext, \hh L}(\vec k, \omega) & = \varepsilon_{\rm eff} \h \vec E_{\rm tot, \hh L}(\vec k, \omega) \,, \\[5pt]
 \vec E_{\rm ext, \hh T}(\vec k, \omega) & = \frac{\varepsilon_{\rm eff} \h \omega^2 - \mu_{\rm eff}^{-1} \h c^2 |\vec k|^2}{\omega^2 - c^2 |\vec k|^2} \, \vec E_{\rm tot, \hh T}(\vec k, \omega) \,.
\end{align}
In particular, this shows that $\varepsilon_0 \hh \vec E_{\rm ext}$ {\itshape does not coincide with the displacement field $\vec D$ used in the phenomenological derivation.}
Instead, the Fundamental Field Identification holds only for the respective longitudinal parts, such that the transverse displacement field $\vec D_{\rm T}$ remains completely
undetermined. In principle, it would then also be unclear how the corresponding transverse response function can actually be measured.
In practice, however, this does not pose any problems since in the microscopic approach, all field quantities are uniquely determined by their respective Maxwell equations.
Correspondingly, {\itshape we here consider Eqs.~\eqref{eq_permitt}--\eqref{eq_permea} as a heuristic ansatz, whose sole purpose lies in the deduction
of the proper fundamental response tensor defined by Eq.~\eqref{proper_fund}.} The interpretation of the material parameters appearing there as ``effective'' permittivity and permeability
can be justified {\itshape ex post,} i.e., independently of the originally macroscopic ansatz. For the electric case, this has already be shown by the above arguments.
It remains  to prove the analogous result for the magnetic material parameter.

Thus, let us next investigate the magnetic properties of the model defined by Eq.~\eqref{proper_fund}. We first note that
the {\itshape direct} fundamental response tensor \cite[\S\,2.3]{Refr} has again the isotropic form \eqref{eq_IsoForm},
such that the longitudinal and transverse components can be calculated as~\cite[\S\,5.1]{EDWave}
\begin{align}
 \chi_{\rm L}(\vec k, \omega) & = \frac{\widetilde \chi_{\rm L}(\vec k, \omega)}{\varepsilon_{\rm L}(\vec k, \omega)} \,, \\[3pt]
\chi_{\rm T}(\vec k, \omega) & = \frac{\widetilde \chi_{\rm T}(\vec k, \omega)}{\varepsilon_{\rm T}(\vec k, \omega)} \,.
\end{align}
From our previous results, we obtain
\begin{align}
\chi_{\rm L}(\vec k, \omega) & = \varepsilon_0 \, \omega^2 \h \bigg( 1 - \frac{1}{\varepsilon_{\rm eff}}
\bigg) \,, \label{fundL} \\[5pt]
\chi_{\rm T}(\vec k, \omega) & = \varepsilon_0 \h (\omega^2 - c^2 |\vec k|^2) \nonumber \\[1pt]
& \quad \, \times \bigg( 1 - \frac{\omega^2 - c^2 |\vec k|^2}{\varepsilon_{\rm eff} \h \omega^2 - \mu_{\rm eff}^{-1} \h c^2 |\vec k|^2} \bigg) \,.\label{fundT}
\end{align}
Furthermore, with the Green function of the d'Alembert operator given by \cite[Eq.~(3.9)]{ED1}
\begin{equation}
\mathbbmsl D_0(\vec k, \omega) = \frac{1}{\varepsilon_0 \h (c^2 |\vec k|^2 - \omega^2)} \,,
\end{equation}
we can write the transverse current response function as
\begin{equation} \label{fundTalt}
 \chi_{\rm T}(\vec k, \omega) = \mathbbmsl D_0^{-1}(\vec k, \omega) \h 
 \bigg( \frac{\omega^2 - c^2 |\vec k|^2}{\varepsilon_{\rm eff} \h \omega^2 - \mu_{\rm eff}^{-1} \h c^2 |\vec k|^2} - 1 \bigg) \,.
\end{equation}
In particular, we note that the density response function (see Eq. \eqref{def_density_response}) is determined by the longitudinal current response function as \cite[Eq.~(7.11)]{ED1}
\begin{equation}
 \upchi(\vec k, \omega) = -\frac{|\vec k|^2}{\omega^2} \h \chi_{\rm L}(\vec k, \omega) \,.
\end{equation}
From Eq.~\eqref{fundL}, we therefore obtain
\begin{equation}
  \upchi(\vec k, \omega) = \varepsilon_0 \h \bigg( \frac{1}{\varepsilon_{\rm eff}} - 1 \bigg) \h |\vec k|^2  \,.
\end{equation}
Finally, the {\itshape magnetic susceptibility} is determined by the transverse current response function as \cite[Eq.~(7.9)]{ED1}
\begin{equation}
 \chi_{\rm m}(\vec k, \omega) = \mathbbmsl D_0(\vec k, \omega) \h \chi_{\rm T}(\vec k, \omega) \,.
\end{equation}
From Eq.~\eqref{fundTalt}, we directly obtain
\begin{equation} \label{zw_1}
 \chi_{\rm m}(\vec k, \omega) = \frac{\omega^2 - c^2 |\vec k|^2}{\varepsilon_{\rm eff} \h \omega^2 - \mu_{\rm eff}^{-1} \h c^2 |\vec k|^2} - 1 \,.
\end{equation}
In particular, the static susceptibility is given by
\begin{equation}
 \chi_{\rm m}(\vec k, \omega = 0) \h = \h \mu_{\rm eff} - 1 \,,
\end{equation}
and this shows that the material constant $\mu_{\rm eff}$ indeed has the interpretation of an effective permeability.
Finally, \linebreak

\pagebreak \noindent
we remark that Eq.~\eqref{zw_1} can also be derived directly from Eq.~\eqref{epsT} by using the general identity \cite[Eq.~(6.48)]{ED1}
\begin{equation}
 \chi_{\rm m}(\vec k, \omega) =  \mu(\vec k, \omega) - 1 \,,
\end{equation}
together with the Universal Response Relation \cite[Eq.~(3.61)]{Refr}
\begin{equation}
 \mu(\vec k, \omega) = \frac{1}{\varepsilon_{\rm T}(\vec k, \omega)}
\end{equation}
between transverse response functions.

\section{Refractive index}

On the microscopic level, the fundamental wave equation for transverse electromagnetic oscillations in (isotropic) materials is given in terms of the transverse dielectric function as \cite{Refr}
\begin{equation}
 \left(-\frac{\omega^2}{c^2} + |\vec k|^2 \right)\varepsilon_{\rm T}(\vec k, \omega) \h \vec E(\vec k,\omega) = 0 \,. \label{eq_MicrWave}
\end{equation}
The standard wave equation of {\itshape ab initio} materials physics,
\begin{equation}
 \left(-\frac{\omega^2}{c^2}\varepsilon_{\rm L}(\vec k,\omega) + |\vec k|^2\right)\vec E(\vec k,\omega) = 0 \,,
\end{equation}
can be obtained from this under the usual assumption of {\it coinciding longitudinal and transverse conductivities} (see Ref.~\cite[\S\,4.4]{Refr}), i.e.,
$\widetilde \sigma_{\rm L}(\vec k, \omega)=\widetilde \sigma_{\rm T}(\vec k, \omega)$.
However, from Eqs.~\eqref{propL}--\eqref{propT} and from the Universal Response Relation $\widetilde \chi(\vec k, \omega)=\i\omega\h\widetilde \sigma(\vec k, \omega)$ between
the current response tensor and the conductivity tensor \cite[\S\,3.2.3]{ED2}, we conclude that this assumption is not fulfilled in our case, 
and hence we have to work directly with the fundamental, microscopic wave equation \eqref{eq_MicrWave}. We first note that in the vacuum case where $\varepsilon=1$, Eq.~\eqref{eq_MicrWave} would revert to the free wave equation (in Fourier space).
In materials, however, we have $\omega\neq c|\vec k|$ such that the pre\-{}factor in brackets can be canceled.
Thus, we obtain the dispersion relation $\omega_{\vec k}$ for the propagation of light in the material from the condition \cite{Refr,EDLor,Dolgov}
\begin{equation}
 \varepsilon_{\rm T}(\vec k, \omega_{\vec k}) = 0 \,.
\end{equation}
Using our result \eqref{epsT}, this yields
\begin{equation}
 \omega_{\vec k}^2 = \frac{1}{\varepsilon_{\rm eff} \hh \mu_{\rm eff}} \, c^2 |\vec k|^2 \,.
\end{equation}
Furthermore, the speed of light is given by the phase velocity,
\begin{equation}
 u_{\vec k} = \frac{\omega_{\vec k}}{|\vec k|} = \frac{c}{\sqrt{\varepsilon_{\rm eff} \hh \mu_{\rm eff}}} \,,
\end{equation}
and this implies for the refractive index, $n_{\vec k}=c/u_{\vec k}$\hh, the standard relation
\begin{equation}
 n^2  = \varepsilon_{\rm eff} \hh \mu_{\rm eff} \,.
\end{equation}
In particular, this implies that the refractive index of our model is {\itshape wavevector independent}.
We have thus shown that the standard formula 
for the refractive index is indeed recovered in this phenomenological model, but in terms of the {\it effective} permeability and permittivity.
By contrast, the same does not hold true for the corresponding microscopic response functions.

\section*{Conclusion}

We have derived a simple, phenomenological model for the microscopic current response tensor which 
corresponds to the macroscopic description of media in terms of
``effective'' permittivity and permeability constants. In particular, we have shown
that the microscopic wave equation in media, $\varepsilon_{\rm r, \hh T}(\vec k,\omega)=0$, which is
formulated in terms of the transverse, frequency- and wavevector-dependent dielectric function, yields back the standard equation $n^2=\varepsilon_{\rm eff}\hh \mu_{\rm eff}$ for the refractive index in terms of these
effective material constants, but not in terms of the corresponding microscopic response functions, i.e.,
 $n^2\neq\varepsilon_{\rm r} \hh \mu_{\rm r}$\hh. Since the microscopic current response tensor is
in principle accessible from {\itshape ab initio} calculations, one could check whether for certain materials it reverts in some appropriate limit to the form \eqref{proper_fund}
with simultaneously negative material constants (as they are expected for metamaterials on grounds of V.~Veselago's argument \cite{Veselago}). Thus, our work also provides a criterion 
for the {\itshape ab initio screening}, i.e., the systematic search for new metamaterials based on modern first-principles calculations.

\section*{Acknowledgments}

This research was supported by the DFG grant HO 2422/12-1 and by the DFG RTG 1995. R.\,S.~thanks the Institute for Theoretical Phys\-{}ics at TU Bergakademie Freiberg for its hospitality. The authors are grateful to Prof.~em.~Franz J.~Wegner (Universit\"at Heidelberg) for illuminating discussions and critical remarks.

\bibliography{/net/home/lxtsfs1/tpc/schober/Ronald/masterbib}

\end{document}